\documentclass[runningheads]{llncs}

\usepackage[T1]{fontenc}
\usepackage{graphicx}
\usepackage{amsmath,amssymb,amsfonts,latexsym}
\usepackage{enumerate}
\usepackage{xspace}
\usepackage{epsf,picinpar}
\usepackage{varioref}
\usepackage{colortbl,multirow,hhline}
\usepackage{listings}
\usepackage{amssymb}
\usepackage{colortbl,multirow,hhline}
\usepackage{algorithmic}
\usepackage{algorithm}
\usepackage{caption}
\usepackage[normalem]{ulem}
\usepackage{xcolor}
\usepackage{pifont}
\usepackage{xcolor,colortbl}
\usepackage{url}
\usepackage{balance}
\usepackage{graphicx, subfigure}
\usepackage{longtable}
\usepackage{lscape}
\usepackage{multirow}
\usepackage{framed}
\usepackage{morefloats}
\usepackage[T1]{fontenc}
\usepackage{array}
\usepackage{pdfpages}
\usepackage{fancybox}
\usepackage{amsmath}
\usepackage{flushend}
\usepackage{booktabs}
\usepackage{enumitem}
\usepackage{tabularx}
\usepackage{makecell}
\usepackage{siunitx}
\usepackage{dirtytalk}
\usepackage{tcolorbox}
\usepackage{ifthen}
\usepackage{csquotes}
\usepackage{textcomp}
\usepackage{soul}
\usepackage{float}
\usepackage[section]{placeins}

\newcommand*{\ie}{i.e.,\@\xspace}

\newcommand{\chg}[1]{\textcolor{black}{#1}}

\newcommand{\energystaticFeaturesCount}{33 }
\newcommand{\energytotalJavaFiles}{768 }
\newcommand{\energytotalJavaMethods}{2,786 }

\newcommand{\totalExecutionTimes}{1,980 }
\newcommand{\totalEnergyValues}{1,103 }

\newcommand{\ROSETTATasksinGitHub}{1,228 }
\newcommand{\ROSETTATasksinGitHubJava}{1,146 }
\newcommand{\ROSETTAJavaFiles}{1,790 }
\newcommand{\RosettaTasksAfterDrop}{805 }
\newcommand{\RosettaJavaAfterDrop}{822 }

\newcommand{\CLBGTasks}{10 }
\newcommand{\CLBGJavaImplementations}{49 }
\newcommand{\CLBGTasksAfterDrop}{7 }
\newcommand{\CLBGJavaAfterDrop}{31 }

\newcommand{\Sum} [2] {\the\numexpr #1 + #2 \relax}

\newcommand{\totalTasks}{\Sum{\CLBGTasks}{\RosettaTasksAfterDrop} }

\newcommand{\rqone}{
  To what extent can method-level energy consumption be predicted from static source code features, \chg{and how does predictive performance change when execution time is added?}
}

\newcommand{\rqtwo}{
  What is the effect of feature selection on the performance of energy prediction models?
}

\newcommand{\rqthree}{
  How does hyperparameter tuning influence model performance for predicting method-level energy usage?
}

\newcommand{\eqMAE}{MAE = \frac{1}{n} \sum_{i=1}^{n} \left| y_i - \hat{y}_i \right|}

\newcommand{\eqMSE}{MSE = \frac{1}{n} \sum_{i=1}^{n} {(y_i - \hat{y}_i )}^2}

\newcommand{\eqRTwo}{R^2 = 1 - \frac{\sum_{t=1}^{n} (y_i - \hat{y}_i)^2}{\sum_{t=1}^{n} (y_i - \bar{y}_i)^2}}

\newcommand{\eqMAPE}{MAPE = \frac{100}{n} \sum_{i=1}^{n} \left| \frac{y_i - \hat{y}_i}{y_i} \right|}

\newboolean{showcomments}
\setboolean{showcomments}{false}

\ifthenelse{\boolean{showcomments}}
{
}

\newboolean{showst}
\setboolean{showst}{false}  

\ifthenelse{\boolean{showst}}
{}  
{\renewcommand{\st}[1]{}}  

\begin{document}
%
\title{Static Metrics Are Insufficient: Predicting Java Method Energy Usage with Execution Time}
\titlerunning{Static Metrics Are Insufficient: Predicting Java Method Energy}

\author{
Muhammad Imran\inst{1} \and
Vincenzo Stoico\inst{2} \and
Ivano Malavolta\inst{2}
}

\institute{
University of L'Aquila, L'Aquila, Italy \\
\email{muhammad.imran@graduate.univaq.it}
\and
Vrije Universiteit Amsterdam, Amsterdam, The Netherlands \\
\email{v.stoico@vu.nl, i.malavolta@vu.nl}
}

\vspace{-1em}
\maketitle
\vspace{-1.5em}

\begin{abstract}
  The increasing energy demand of software systems is raising concerns about their environmental impact and associated costs. Reasoning on energy usage early in the development flow has the potential to significantly reduce the overall energy usage of a software system, \chg{as it allows developers to make informed design and refactoring decisions before inefficiencies propagate}. However, \chg{assessing energy usage without repeated profiling and direct measurement is difficult, which limits early reasoning in practice.} 
  This study investigates the limits of method-level energy prediction in Java, examining whether static source code metrics complemented with method-level execution time can estimate the energy consumption of Java methods. We profile \energytotalJavaMethods Java methods to extract \energystaticFeaturesCount static features and measure execution time and energy, then train and compare eleven regression models. 
  Our findings show that static source code metrics alone yield poor predictive performance, with average $R^2$ values close to zero. Incorporating execution time as a lightweight dynamic input significantly improves accuracy, raising $R^2$ to as high as 0.46. Execution time, internal method calls, and cyclomatic complexity consistently emerge as the strongest predictors of energy consumption.


\end{abstract}

\section{Introduction}
\vspace{-0.6em}
Energy consumption has emerged as a critical concern in modern software development, not only in mobile or embedded environments but also in general-purpose computing. As software systems grow increasingly complex and resource-intensive, their energy footprints become non-trivial, impacting battery life, operational costs, and environmental sustainability \cite{manner2023black}.

A significant part of the research in this domain relies on empirical measurements, as energy is strongly influenced by the execution environment of the application \cite{castor2024estimating}. These measurements are obtained through repeated executions in controlled environments, often using specialized profiling tools and hardware instrumentation \cite{schuler2024systematic}. Although these methods provide precise and reliable energy consumption data, they often require manual setup or hardware access to capture fine-grained data, which limits their routine use in development workflows \cite{georgiou2019software}. Consequently, during refactoring or design, developers typically lack accessible ways to reason about energy consumption without repeated profiling \cite{oliveira2016native}. The limitations of measurement-based approaches become especially clear when energy profiling is required at fine-grained levels, such as individual methods. Recent studies demonstrate that even small syntactic code changes can introduce measurable variations in energy usage, especially in compiled or performance-sensitive programs \cite{oliveira2021improving}.

Despite this growing recognition, few studies have systematically modeled energy consumption from static code features like cyclomatic complexity, loop depth, or library usage, especially at the method level. Multiple studies \cite{grech2015static,alvi2021mlee} show static features influence energy by shaping control flow, computational intensity, and resource access, highlighting their importance in energy modeling. Most work focuses on coarse-grained estimation (language- or library-level) or hybrid methods requiring execution \cite{oliveira2021improving,alvi2021mlee}. This gap is particularly notable in Java, where low-level decisions such as collection types and loop constructs affect energy in subtle ways~\cite{oliveira2021improving,kumar2017energy}, while higher-level design patterns have been found largely energy-neutral~\cite{noureddine2025investigating}. These observations motivate approaches that support fine-grained energy awareness while reducing reliance on repeated profiling and specialized instrumentation, since energy consumption depends not only on source code but also on its interaction with hardware and runtime characteristics~\cite{pereira2017energy}, suggesting that combining static and dynamic features may improve prediction accuracy, \chg{and at method level, such predictions can directly guide developers toward the specific code units responsible for high energy usage, enabling targeted optimization.}


The \textbf{goal} of this study is to assess the extent to which static features extracted from Java methods can explain method-level energy consumption, and to identify their limitations. We further investigate whether adding execution time improves predictive performance. Our research methodology involves profiling \energytotalJavaMethods Java methods from \energytotalJavaFiles files representing \totalTasks programming tasks to collect static features, execution time, and energy measurements. We extract \energystaticFeaturesCount source-level features per method, profile execution time and energy on a fixed testbed, and train eleven regression models through a structured process including feature selection, model comparison, and hyperparameter tuning.








Our findings show that static code features alone yield near-zero predictive power ($R^2 \approx 0$), and that only the inclusion of execution time as a lightweight dynamic input raises $R^2$ to 0.46, indicating that method-level energy is dominated by runtime behavior rather than source structure alone.

We provide a publicly available replication package \cite{replicationpackage_method_energy_java_2025}, containing the dataset, energy measurements, execution times, static features, training scripts, and complete experimental results.

\FloatBarrier
\vspace{-0.6em}
\section{Related Work}\label{sec:related}
\vspace{-0.6em}

\paragraph{\textbf{Energy Efficiency}}
Recent studies have explored energy efficiency in high-level languages, showing that alternative compilers such as Codon~\cite{shajii2023codon}, Pythran, and PyPy consistently outperform CPython in execution time and energy efficiency~\cite{augier2021reducing}, and that HOPE~\cite{akeret2015hope} and Codon demonstrate similar gains on numerical benchmarks against C++. While these studies show energy gains through compiler optimization, they focus on coarse-grained program behavior and specific benchmark types or parallel computing use cases.

Recent work has also examined fine-grained energy variations caused by structural differences in source code. Stoico et al. \cite{stoico2025empirical} demonstrated  that small syntactic modifications in Python programs, such as altering list comprehensions or loop constructs, can significantly affect both execution time and energy consumption, depending on how the code is compiled and optimized. Similarly, Cursaru et al. \cite{cursaru2024controlled} showed that functionally equivalent code generated by different LLM prompts can result in measurable differences in energy usage, driven solely by structural variations. These studies emphasize that source code organization itself can directly influence energy behavior, even though their results are not directly transferable to Java, and motivate the broader question of whether structural source code properties are informative for energy estimation at finer levels of granularity. We investigate this question at the level of individual Java methods using static code features and execution time.

\vspace{-0.3em}
\paragraph{\textbf{Static Code Metrics}}
Several studies have explored static code metrics as predictors of software energy consumption. Kumar et al.~\cite{kumar2017energy} showed that Java constructs such as loops, data types, and collections can significantly impact energy usage. Hamza Alvi et al.~\cite{alvi2021mlee} proposed the MLEE framework, which applies machine learning for method-level energy estimation in Android applications using structural code features under controlled execution scenarios, though its strong predictive performance relies on platform-specific assumptions not directly transferable to general-purpose Java workloads. Akinli~\cite{akinli2018software} further demonstrated the viability of combining code metrics with learning models for energy prediction across benchmarks. Broader perspectives are offered by Kruglov et al.~\cite{kruglov2023developing}, who emphasize the integration of sustainability metrics into software processes, and Schuler et al.~\cite{schuler2024systematic}, who identify the lack of fine-grained static code-based energy models in the current literature. Beyond energy estimation, static metrics have also proven valuable in predicting security vulnerabilities~\cite{ganesh2022source}, estimating maintenance effort~\cite{chowdhury2022revisiting}, and understanding the energy impact of design decisions such as code smells and refactoring~\cite{poy2024impact}. Closest to our work, Goyal et al.~\cite{goyal2026encode} recently proposed EnCoDe, which predicts block-level energy of Python code purely from static AST features ($R^2$ of 0.75), with ground-truth labels obtained via amplified executions of isolated code blocks with fixed inputs. Notably, their ablation shows that complexity metrics alone yield an $R^2$ of 0.067. In contrast, our study examines a broader set of static features including metrics from software performance evaluation, adds execution time as a dynamic feature, \chg{targets methods executed within complete Java programs rather than isolated code fragments,} and performs a structured comparison of multiple regression models with systematic feature selection and hyperparameter tuning, offering a complete evaluation of ML-based energy estimation at the method level.

\paragraph{\textbf{Machine Learning Models for Prediction}}
While MLEE~\cite{alvi2021mlee} and Akinli~\cite{akinli2018software} demonstrated the feasibility of ML-based energy estimation at method level, neither performed extensive model comparison, tuning, or feature selection analysis. Kruglov et al. \cite{kruglov2023developing} and Schuler et al. \cite{schuler2024systematic} also note the growing trend of model-based estimation, yet few studies systematically compare the performance of different models or assess the impact of tuning and feature selection on prediction accuracy. Our study addresses this gap through a structured comparison of eleven regression models evaluated across multiple feature selection and tuning configurations.

\FloatBarrier
\vspace{-0.6em}
\section{Study Design}
\vspace{-0.6em}
We define three research questions (RQs) guiding our investigation:

$\triangleright$ \textbf{RQ$_1$}: \textit{\rqone} This research question assesses the baseline predictive value of static code features and evaluates the marginal effect of adding execution time as a lightweight dynamic input. Models are trained with default hyperparameters.

$\triangleright$ \textbf{RQ$_2$}: \textit{\rqtwo} We investigate whether removing less informative features improves generalization and which code characteristics are most predictive of energy usage, offering guidance for developers when dynamic profiling is unavailable.

$\triangleright$ \textbf{RQ$_3$}: \textit{\rqthree} We explore whether fine-tuning yields meaningful accuracy gains over defaults, clarifying the relative importance of model optimization versus feature engineering in our prediction setting.

\begin{figure}[t!]
  \centering
  \begin{minipage}[t]{0.44\textwidth}
    \centering
    \includegraphics[width=\textwidth,keepaspectratio]{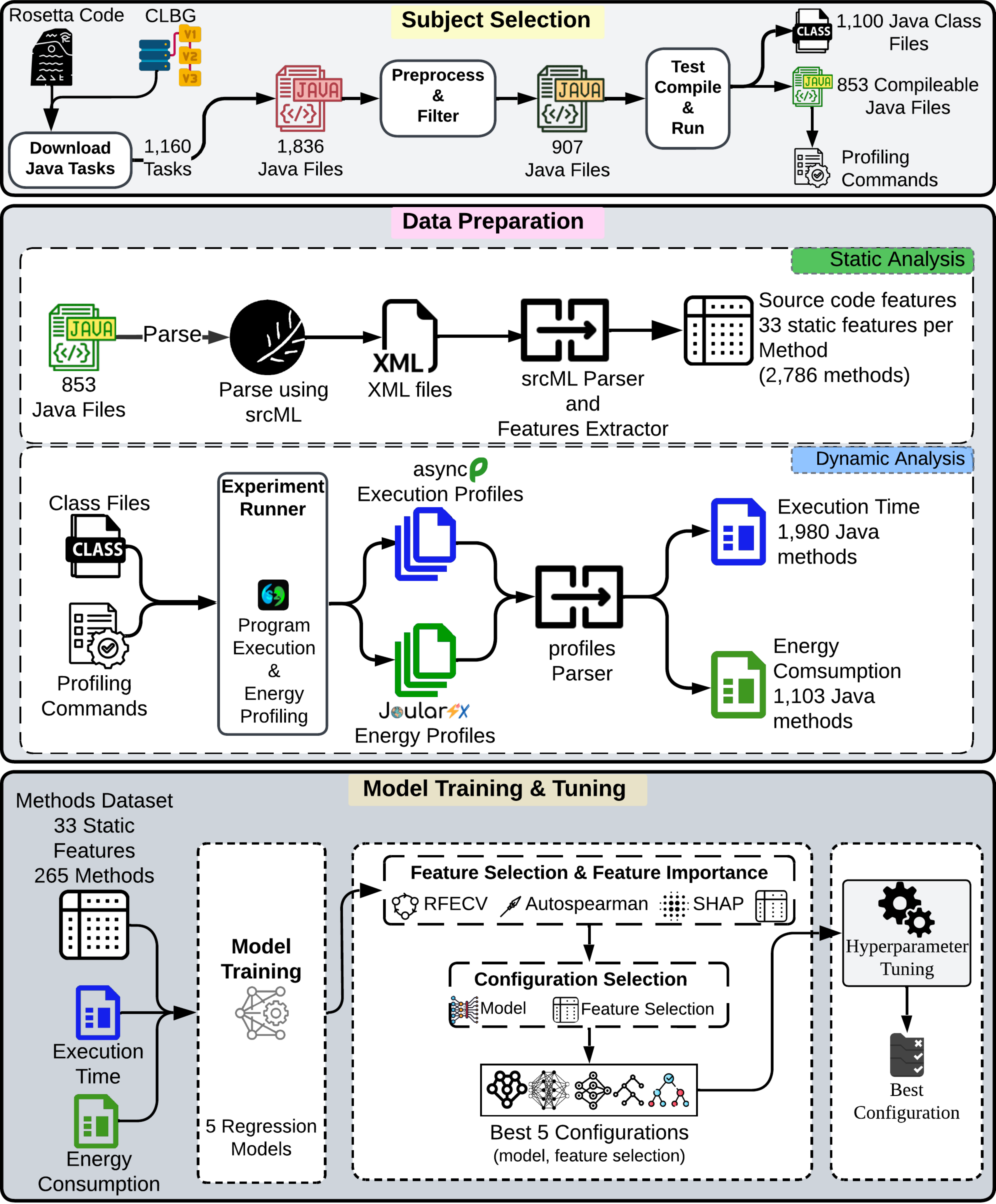}
    \vspace{1mm}
    \textbf{(a)} Main steps of the study.
  \end{minipage}
  \hfill
  \begin{minipage}[t]{0.52\textwidth}
    \centering
    \includegraphics[width=\textwidth,keepaspectratio]{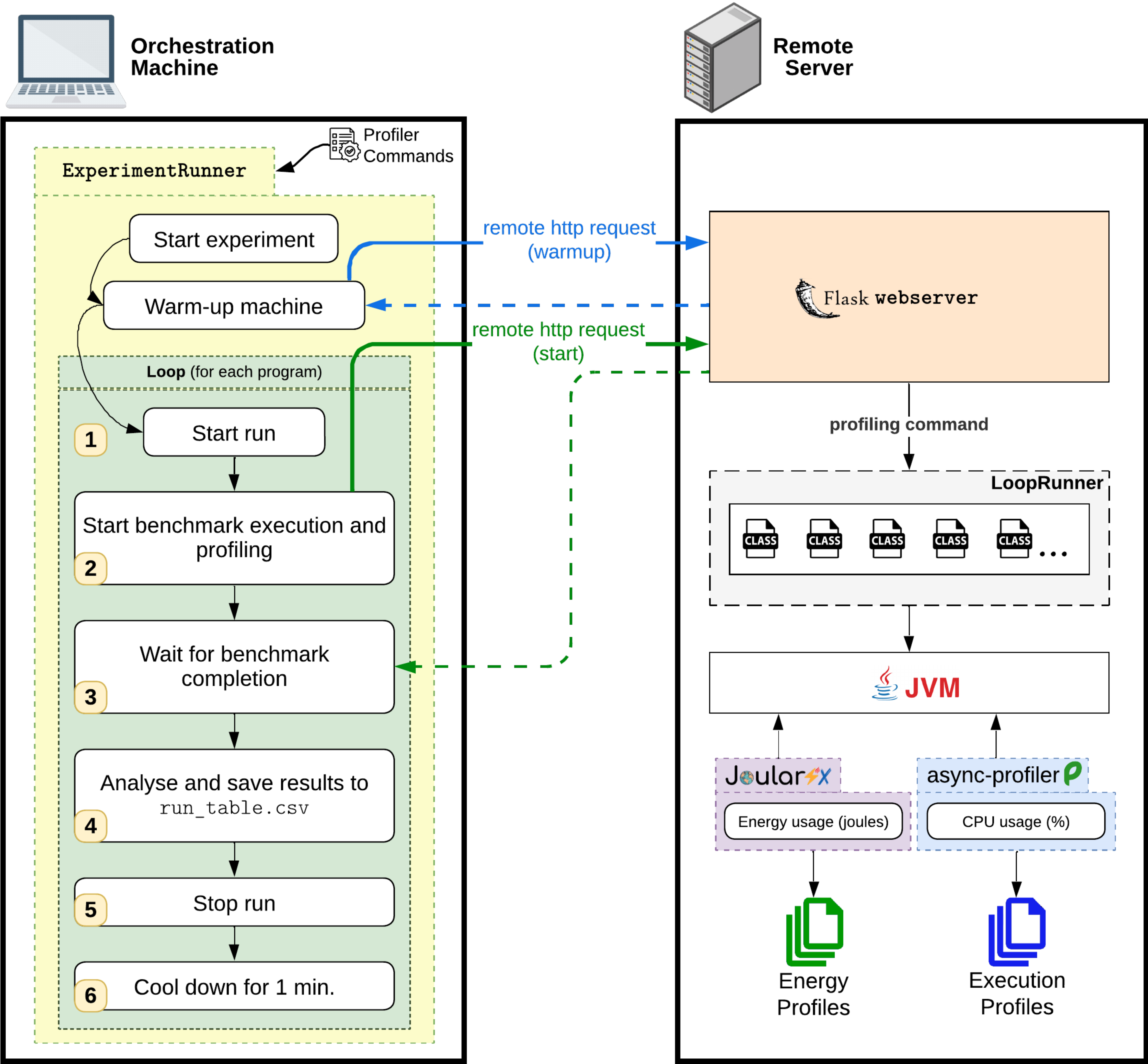}
    \vspace{1mm}
    \textbf{(b)} Profiling using Experiment Runner.
  \end{minipage}
  \caption{Overview of the study workflow and profiling setup.}
  \label{fig:study-profiling}
\end{figure}
Our study follows a structured process consisting of three main steps: \textit{(i)} \emph{subject selection}, \textit{(ii)} \emph{data preparation}, and \textit{(iii)} \emph{model training and tuning}, as shown in Figure~\ref{fig:study-profiling}(a).

\vspace{-0.4em}
\subsection{\textbf{Subject Selection}}
\vspace{-0.4em}

We selected Java implementations from two publicly available repositories: the Computer Language Benchmarks Game (CLBG)\footnote{\url{https://benchmarksgame-team.pages.debian.net/benchmarksgame/index.html}} and Rosetta Code\footnote{\url{https://rosettacode.org/wiki/Rosetta_Code}}. They provide community-contributed implementations of algorithmic tasks across multiple languages, typically self-contained and computationally intensive, making them well suited for controlled energy profiling at method level. For Rosetta Code, we used a curated GitHub snapshot\footnote{\url{https://github.com/acmeism/RosettaCodeData}} organizing \ROSETTATasksinGitHub tasks, of which \ROSETTATasksinGitHubJava had Java implementations, resulting in \ROSETTAJavaFiles files across all variants. From CLBG, crawling \CLBGTasks tasks provided \CLBGJavaImplementations files.

We filtered out implementations that: (i) lacked a \texttt{main} method, (ii) required user input or a graphical interface, (iii) contained infinite loops, (iv) had duplicate class names, or (v) could not be compiled due to external dependencies. After filtering \Sum{\CLBGJavaAfterDrop}{\RosettaJavaAfterDrop} Java files corresponding to \Sum{\CLBGTasksAfterDrop}{\RosettaTasksAfterDrop} distinct tasks were retained.





\vspace{-0.4em}
\subsection{\textbf{Data Preparation}} \label{subsec:data_prep}
\vspace{-0.4em}
Data was collected at method level through two complementary pipelines: static analysis for code features and dynamic profiling for execution time and energy.
\vspace{-0.4em}
\paragraph{\textbf{Static Analysis}}
Following established practices in performance analysis~\cite{imran2024empirical}, each Java file was parsed using srcML~\cite{Collard2011srcML} to apply XPath queries. We extracted \energystaticFeaturesCount static features per method covering control flow, complexity, and standard library usage, as summarized in Table~\ref{tab:energy_static-features}. Features were guided by prior work on software performance~\cite{imran2025code} and selected by correlating known static metrics with energy to filter out less informative ones, yielding features for \energytotalJavaMethods methods across the \Sum{\CLBGJavaAfterDrop}{\RosettaJavaAfterDrop} files.

\vspace{-0.4em}
\paragraph{\textbf{Dynamic Profiling}}
To collect execution time and energy consumption data, we followed established guidelines for energy measurement~\cite{FGCS_2024,EMSE_EDU_2024} and used Experiment Runner \cite{KARSTEN2026103415}
as experiment orchestrator shown in Figure~\ref{fig:study-profiling}(b). For CPU sampling, we used \textit{async-profiler}\footnote{\url{https://github.com/async-profiler/async-profiler}} to collect method-level execution times, and for energy measurement, we employed \textit{JoularJX}\footnote{\url{https://www.noureddine.org/research/joular/joularjx}}, which provides method-level energy estimation within the JVM.

\begin{table}[t!]

  \renewcommand{\arraystretch}{1} 
  \centering

  \caption{Collected source code features.}

  \label{tab:energy_static-features}
  \scriptsize
  \begin{tabularx}{\textwidth}{|c|c|l|X|}
    \hline
     & \textbf{Level/ Category} & \textbf{Feature Name} & \textbf{Description} \\
    \hline

    \multirow{4}{*}{\rotatebox[origin=c]{90}{\makecell[c]{\textbf{(i)meta info}}}}
    & \multirow{4}{*}{\makecell[c]{method}}
    & \rule{0pt}{2.9ex}\#methodLOC & Lines of code of the method \\
    \cline{3-4}
    & & \rule{0pt}{2.9ex}\#nameLen & method name length \\
    \cline{3-4}
    & & \rule{0pt}{2.9ex}methodScope & access specifier of the method \\
    \cline{3-4}
    & & \rule{0pt}{2.9ex}isOverloaded & if the method is overloaded \\
    \hline

    \multirow{15}{*}{\rotatebox[origin=c]{90}{\makecell[c]{\textbf{(ii) language feature}}}}
    & \multirow{15}{*}{\shortstack[c]{control flow\\and data}}
    & \#if & no. of if conditions \\

    \cline{3-4} & & \#switch & no. of switch statements \\
    \cline{3-4} & & \#case & no. of case statements \\
    \cline{3-4} & & \#for & no. of for loops \\
    \cline{3-4} & & \#while & no. of while loops \\
    \cline{3-4} & & \#do & no. of doWhile loops \\
    \cline{3-4} & & \#nestedLoops & no. of nested loops (arbitrary depth) \\
    \cline{3-4} & & \#methodCalls & no. of methods called \\
    \cline{3-4} & & \#internalCalls & no. of internal methods called \\
    \cline{3-4} & & \#externalCalls & no. of external methods called \\
    \cline{3-4} & & \#return & no. of return statements \\
    \cline{3-4} & & \#throw & no. of throw statements \\
    \cline{3-4} & & \#catch & no. of catch statements \\
    \cline{3-4} & & \#cyclo & cyclomatic complexity \\
    \cline{3-4} & & \#vars & no. of the variables declared \\

    \hline

    \multirow{15}{*}{\rotatebox[origin=c]{90}{\makecell[c]{\textbf{(iii) standard Java APIs}}}}
    & utility
    & java.util & Utility classes in Java. \\

    \cline{2-4}
    & \multirow{2}{*}{os and concurrency}
    & java.lang & Core Java classes \& Multithreading \\
    \cline{3-4}
    & & java.lang.management & Management interfaces for Java. \\
    \cline{3-4}
    & & java.util.concurrent & Advanced concurrency utilities. \\

    \cline{2-4}
    & \multirow{7}{*}{io}
    & java.io & Data streams based I/O. \\
    \cline{3-4}
    & & java.nio & New I/O for scalable I/O operations. \\
    \cline{3-4}
    & & java.nio.channels & Channels for non blocking I/O. \\
    \cline{3-4}
    & & java.nio.file & NIO based File I/O enhancements. \\
    \cline{3-4}
    & & java.nio.charset & Classes for encoding and decoding. \\
    \cline{3-4}
    & & java.net & Networking and communication. \\
    \cline{3-4}
    & & javax.net.ssl & For secure network communication. \\


    \cline{2-4}
    & \multirow{2}{*}{strings}
    & java.util.regex & Regex for pattern matching \& Strings. \\
    \cline{3-4}
    & & java.text & Text parsing and formatting classes. \\

    \cline{2-4}
    & math
    & java.math & Mathematical utilities. \\

    \hline
  \end{tabularx}
\end{table}

The \Sum{\CLBGJavaAfterDrop}{\RosettaJavaAfterDrop} Java files were compiled using \textit{javac} version \textit{18.0.1.1}, producing 1,100 \textit{.class} files. Profiling commands attach both agents to the JVM via a custom wrapper class, \textit{LoopRunner}, which executes each program's \textit{main} method multiple times in a loop. Since both \textit{async-profiler} and \textit{JoularJX} rely on sampling, repeating executions increases the likelihood of capturing short-lived methods that may not be observed in a single run. We set the repetition count to 20, which captured energy measurements for $\sim$48\% more methods and execution-time data for $\sim$14\% more methods compared to a single run. Input values for CLBG tasks were taken from the task pages and Rosetta Code inputs were embedded in the implementations.


We selected \textit{JoularJX} as the energy profiler because it provides method-level energy measurements directly within the JVM through agent-based instrumentation, correlating runtime method execution with CPU utilization and power readings to attribute total JVM energy consumption to individual methods~\cite{noureddine-ie-2022}. Other approaches, including hardware counters such as Intel's RAPL~\cite{kifetew2025energy} and tools such as CodeCarbon~\cite{lacoste2019quantifying}, PowerAPI~\cite{fieni2024powerapi}, and RJoules~\cite{chattaraj2023rjoules}, produce system or process level estimates rather than method-level attribution. \textit{JoularJX} instead remains lightweight, cross-platform, and compatible with \textit{async-profiler} for synchronized CPU and energy traces. Prior to profiling, a warm-up phase (repeated \texttt{Fibonacci} execution) stabilized JVM state (JIT activation and thermal steady state). Each profiling iteration was triggered via \texttt{HTTP POST} to a Flask server on the testbed, followed by a 30-second cool-down period as a standard practice\footnote{\url{https://luiscruz.github.io/2021/10/10/scientific-guide.html}} for preventing heat accumulation across sequential runs. Both profilers were configured at a \SI{1}{\milli\second} sampling interval to match JoularJX's minimum resolution\footnote{\url{github.com/joular/joularjx/blob/develop/config.properties}}.

Profiler outputs (\textit{.collapsed} for \textit{async-profiler}, \textit{.csv} for \textit{JoularJX}) were parsed to extract method-level metrics. Execution times were recovered for \totalExecutionTimes methods and energy values for \totalEnergyValues methods. Since \textit{async-profiler} samples via CPU events while \textit{JoularJX} operates at fixed temporal intervals~\cite{burchell2023don}, profiler coverage is only partially overlapping, reducing the number of methods with complete feature pairs without affecting measurement validity. To mitigate bias, we retained only the intersection of methods reported by both profilers for model training.

\vspace{-0.6em}
\subsection{\textbf{Model Training and Tuning}}
\vspace{-0.6em}
Our approach was inspired by previous studies on software performance \cite{imran2025code,laaber2021predicting} and energy estimation \cite{akinli2018software}, using regression models to predict energy from source code characteristics. Table~\ref{tab:models_employed_energy} lists the eleven regression models selected based on: (i) prior use in software engineering analytics~\cite{imran2025code,laaber2021predicting,akinli2018software}, (ii) flexibility to capture linear and non-linear relationships, and (iii) availability in scikit-learn\footnote{\url{https://scikit-learn.org/}} for reproducibility. The process follows three steps aligned with RQ1--RQ3: base model training, feature selection, and hyperparameter tuning.

\vspace{-0.4em}
\FloatBarrier

\begin{table}[htbp]
    \centering
    \caption{ML models used in this study.}
    \label{tab:models_employed_energy}
    \footnotesize
    \begin{tabular}{|l|l|l|}
      \hline
      \textbf{Model} & \textbf{Acronym} & \textbf{Algorithm} \\ \hline
      Random Forest & RF & Ensemble \\ \hline
      Gradient Boosting & GB & Ensemble-Boosting \\ \hline
      ADA Boosting & ADA & Ensemble-Boosting \\ \hline
      Hist Gradient Boosting & HGB & Ensemble-Boosting \\ \hline
      Lasso Regression & Lasso-R & Linear \\ \hline
      k-Nearest Neighbor & kNN & Instance-based \\ \hline
      Ridge Regression & Ridge-R & Linear \\ \hline
      Support Vector Machine & SVM & Kernel-based \\ \hline
      Linear Regression & LR & Linear \\ \hline
      Multi-Layer Perceptron & MLP & Neural Network \\ \hline
      Decision Tree & DT & Decision Tree \\ \hline
    \end{tabular}
\end{table}

\vspace{-0.4em}

\paragraph{Data Pre-processing}
From the initial \energytotalJavaMethods methods, matching with profiler outputs (JoularJX and async-profiler) resulted in 902 methods with complete static and dynamic data. After removing methods with zero energy consumption which are likely trivial code paths that would distort training, we got 265 methods remaining. To deal with the long-tailed distributions of both energy and execution time values, we applied a natural logarithmic transformation, which mitigated skewness and made the variables more suitable for regression modeling. The categorical \textit{methodScope} feature was one-hot encoded to produce a fully numeric feature matrix.

\paragraph{Model Training}
We define a supervised learning task where each instance corresponds to a Java method characterized by static features (e.g., lines of code, number of loops), and the target variable is the dynamic energy measure obtained through our profiling setup. The training set includes static features and execution time in milliseconds as predictors and energy consumption (measured in Joules) as the target variable. We apply 5-fold cross-validation (rather than 10-fold, to ensure sufficiently large validation samples given our dataset size of 265 methods) to reduce variance due to sampling, using an 80/20 split repeated five times across shuffled partitions. Models are trained with default hyperparameters without feature pre-processing, as this initial step establishes the baseline performance evaluated in $RQ1$. 

\vspace{-0.4em}
\paragraph{Feature Selection and Configuration Comparison}
To address $RQ2$, we applied seven feature pre-processing techniques: RFECV~\cite{guyon2002gene}, AutoSpearman~\cite{jiarpakdee2018autospearman}, SelectKBest~\cite{scikit-learn-selectkbest} ($k \in \{10, 20, 30\}$), and VarianceThreshold~\cite{sklearn-variancethreshold}. RFECV eliminates features iteratively based on model feedback; AutoSpearman removes multicollinear features by thresholding pairwise Spearman correlation. Evaluating all model $\times$ feature-selection combinations yields 77 configurations (11 models $\times$ 7 variants). Configurations are compared using $R^2$ and the five top-performing configurations are carried forward for hyperparameter tuning.

\vspace{-0.4em}
\paragraph{Hyperparameter Tuning}
To address $RQ3$, we applied \textit{RandomizedSearchCV} to the top five configurations using model-specific parameter grids available 
in the replication package~\cite{replicationpackage_method_energy_java_2025}. 
Random search is more efficient than grid or manual search for finding optimal 
hyperparameters~\cite{bergstra2012random}. The tuning objective is to maximize $R^2$.

\vspace{-0.4em}
\paragraph{Performance Evaluation}
We adopt four standard regression metrics
, widely used in software effort estimation~\cite{bener2015lessons}, defect prediction~\cite{baskeles2007software}, and energy-aware analytics~\cite{akinli2018software,kruglov2023developing,alvi2021mlee}. We use $R^2$ as the primary indicator of overall performance and for selecting the best model configurations, as it captures how effectively the features explain variance in energy usage. Mean Absolute Error (MAE), Mean Squared Error (MSE), and Mean Absolute Percentage Error (MAPE) complement this by characterizing the magnitude and scale of prediction errors.

\FloatBarrier
\vspace{-0.6em}
\section{Results}
\vspace{-0.6em}
\subsection{RQ$_1$: Predictive performance for energy}

\begin{figure}[htbp]
  \centering
  \includegraphics[width=0.60\textwidth]{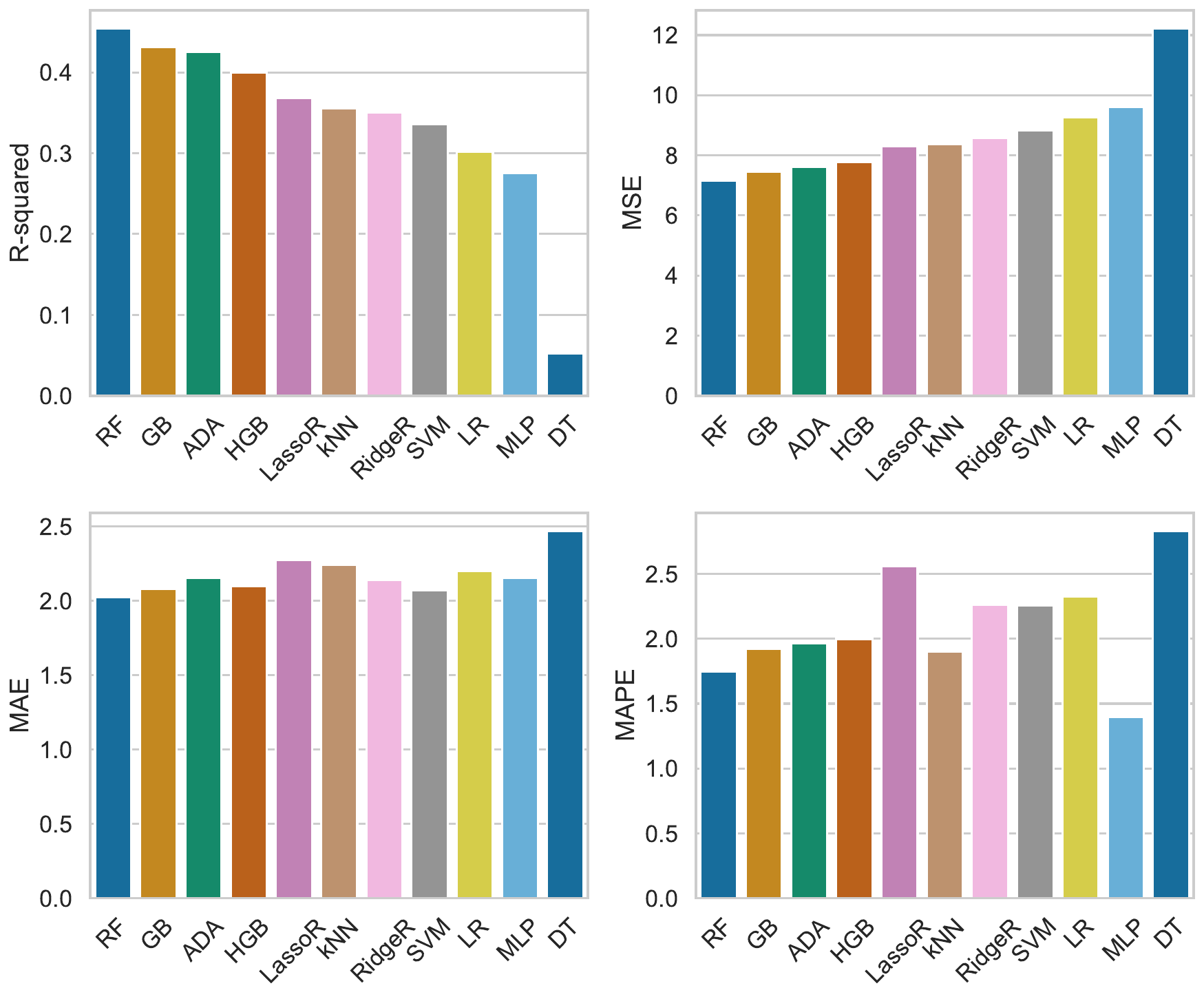}
  \caption{Baseline prediction performance.}
  \label{fig:rq1-results}
\end{figure}

Figure~\ref{fig:rq1-results} summarizes the baseline results, showing that the predictive performance of all regression models trained on static features and execution time remains limited. \texttt{RF} achieves the best performance with an $R^2$ of 0.45, an $MSE$ of 7.15, an $MAE$ of 2.02, and a $MAPE$ of 1.75, offering strong interpretability through its feature importance estimates. At the other extreme, \texttt{DT} performs worst with an $R^2$ of only 0.05 and a $MAPE$ exceeding 2.8, indicating overfitting and poor generalization. Overall, all models only partially explain the variability in energy consumption: the best $R^2$ remains below 0.5, $MSE$ ranges from 7 to 12, and $MAPE$ stays mostly below 2\%, indicating that predictions are proportionally close to actual values even when absolute errors are larger.


\subsection{RQ$_2$: 
\vspace{-0.4em}
Feature selection}
\begin{figure}[htbp]
  \centering
  \includegraphics[width=0.70\textwidth]{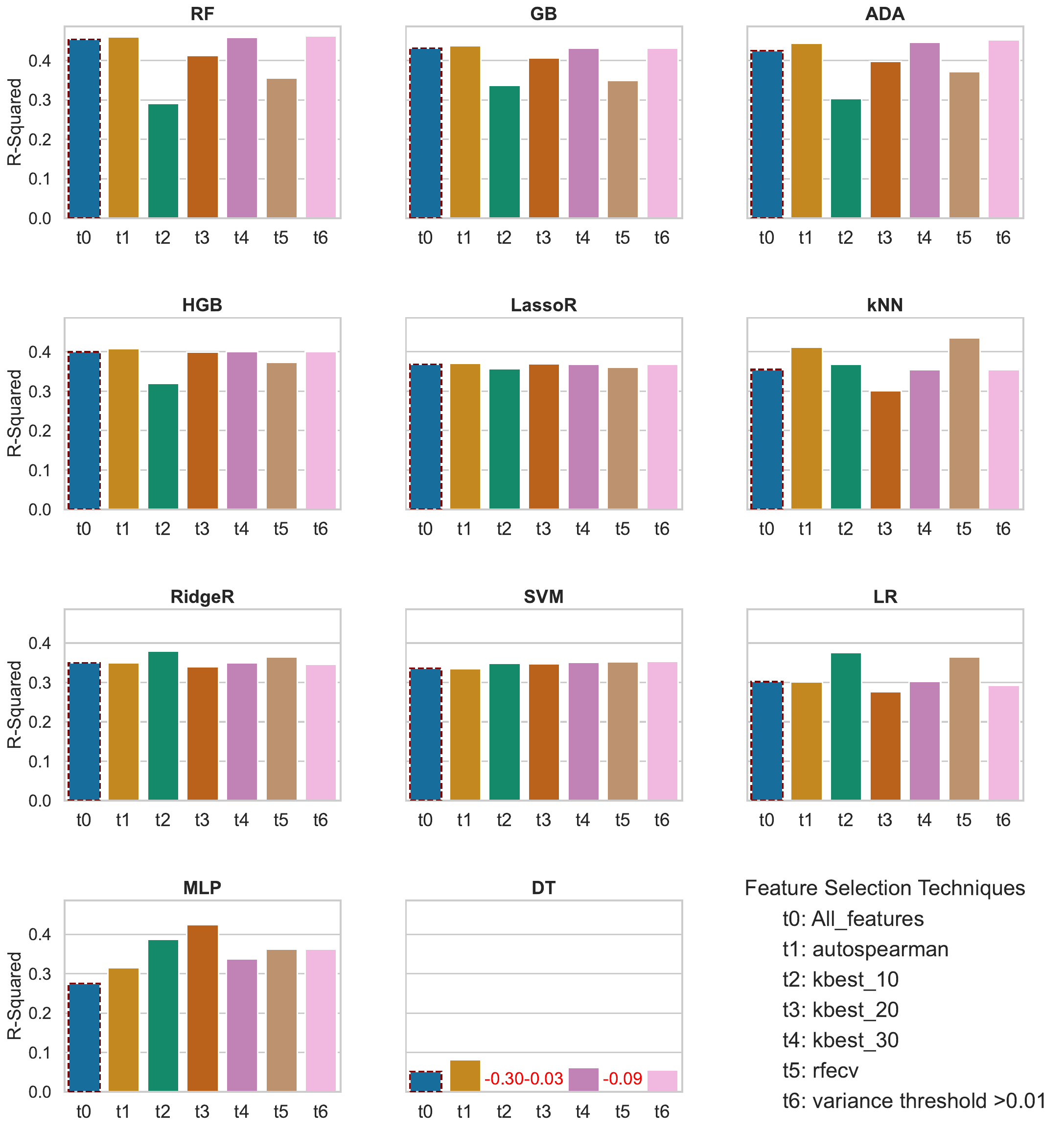}
  \caption{Feature-selection performance.}
  \label{fig:rq2-results}
\end{figure}

We evaluated six feature preprocessing techniques against the baseline using all features. Figure~\ref{fig:rq2-results} reports the $R^2$ values for all 77 configurations, covering 11 models and seven feature settings. The results show that performance varies across both models and preprocessing techniques, although the gains over the baseline are generally small.

Table~\ref{tab:top5_regression_noHyper} reports the five best configurations. RF is the most consistent model, appearing in four of the top five configurations. The best result is obtained with Variance Threshold (t6), with an $R^2$ of 0.463, followed closely by AutoSpearman (0.460), SelectKBest with $k=30$ (0.459), and the full feature set (0.454). The only non RF configuration in the top five is AdaBoost with Variance Threshold, reaching an $R^2$ of 0.453.

\renewcommand{\theadgape}{\Gape[0.5pt][0.5pt]}

\begin{table}[h!]
  \scriptsize
  \centering
  \caption{\textcolor{black}{Top 5 configurations based on $R^2$ score without hyperparameter tuning}}

  \setlength{\tabcolsep}{0pt}
  \begin{tabular*}{0.95\linewidth}{@{\extracolsep{\fill}}l*{6}{r}}
    \toprule

    {\thead[l]{Model}}
    & {\thead[r]{Feature\\ Selector}}
    & {\thead[r]{Features\\ Selected}}
    & {\thead[r]{MSE}}
    & {\thead[r]{MAE}}
    & {\thead[r]{MAPE}}
    & {\thead[r]{$R^2$}}\\
    \midrule

    RF  & Variance Threshold       & 24 & 7.038 & 2.008 & 1.693 & 0.463 \\
    RF  & AutoSpearman             & 22 & 7.058 & 1.998 & 1.873 & 0.460 \\
    RF  & SelectKBest              & 30 & 7.102 & 2.015 & 1.742 & 0.459 \\
    RF  & None                     & 35 & 7.152 & 2.024 & 1.745 & 0.454 \\
    ADA & Variance Threshold       & 24 & 7.218 & 2.162 & 2.152 & 0.453 \\

    \bottomrule
  \end{tabular*}
  \label{tab:top5_regression_noHyper}
\end{table}

\begin{figure}[b]
  \centering
  \begin{minipage}[t]{0.54\textwidth}
    \centering
    \includegraphics[width=\textwidth]{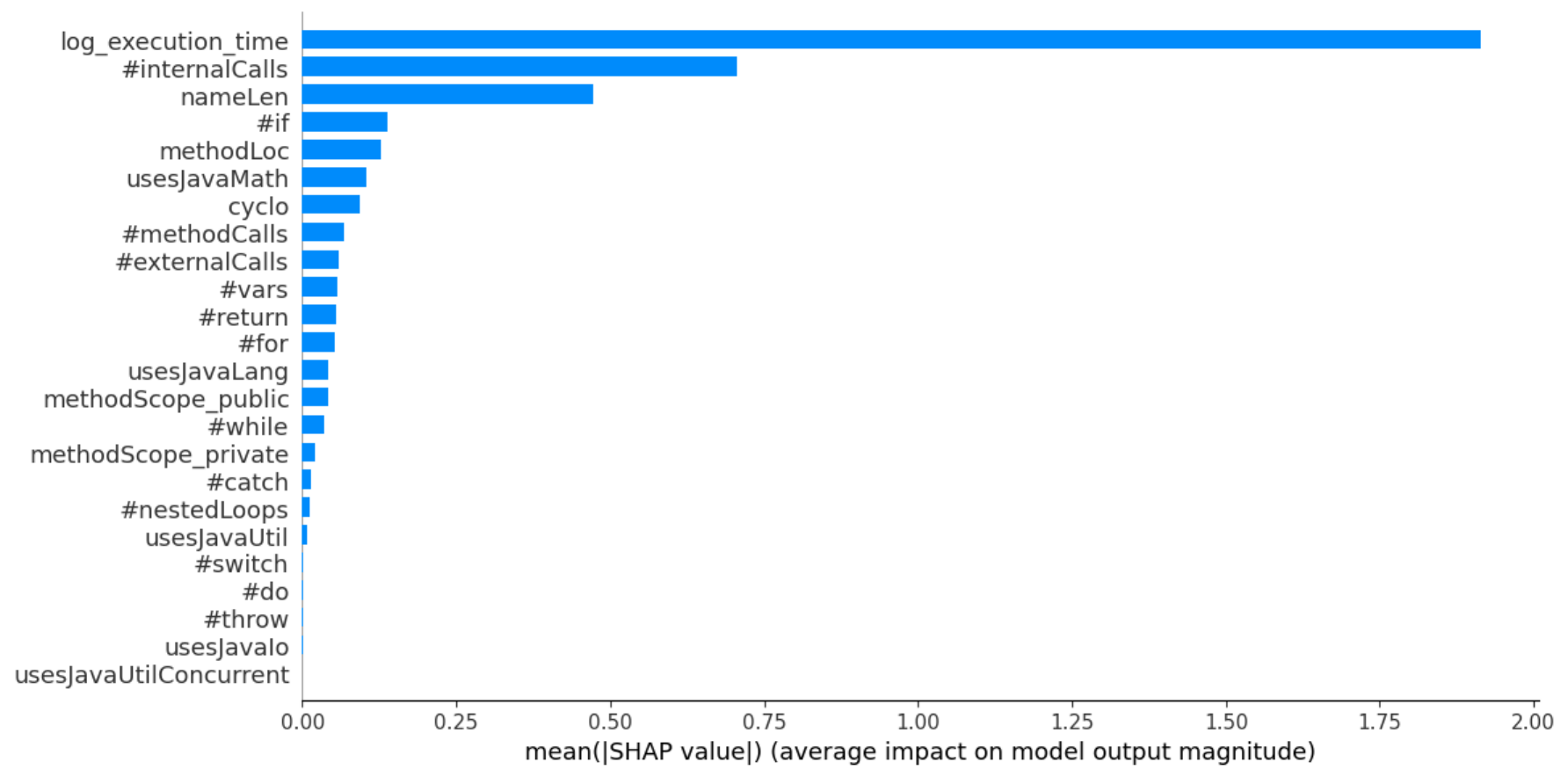}
    \vspace{1mm}
    \textbf{(a)} SHAP values for RF with 24 features selected by Variance Threshold (t6).
  \end{minipage}
  \hfill
  \begin{minipage}[t]{0.45\textwidth}
    \centering
    \includegraphics[width=\textwidth]{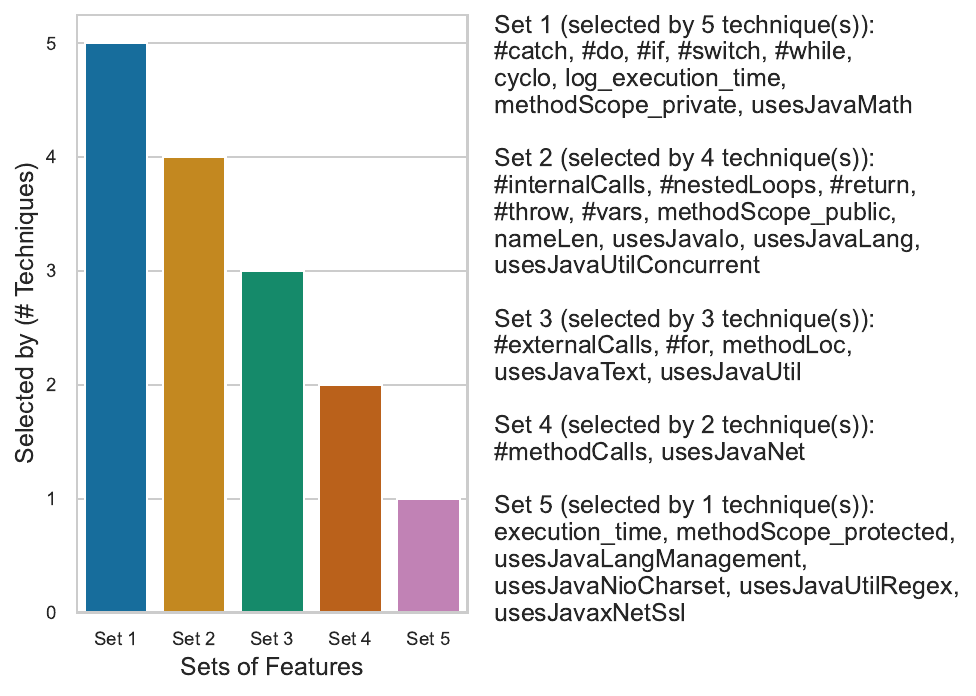}
    \vspace{1mm}
    \textbf{(b)} Frequency of features selected by different techniques.
  \end{minipage}
  \caption{Feature importance analysis for RQ$_2$}
  \label{fig:rq2-feature-analysis}
\end{figure}

Figure~\ref{fig:rq2-feature-analysis}(a) shows the SHAP values for the best configuration. \textit{log\_execution\_time} is the most influential feature, followed by \textit{\#internalCalls} and other structural features related to control flow and method behavior. Figure~\ref{fig:rq2-feature-analysis}(b) shows how often each feature was selected. Features such as \textit{\#catch}, \textit{\#do}, \textit{\#cyclo}, and \textit{log\_execution\_time} are selected by multiple techniques, while features such as \textit{usesJavaUtilConcurrent} and \textit{usesJavaLangManagement} appear less consistently, suggesting that a core subset of structural features contributes more reliably to energy estimation. One non-obvious feature among the top ranks is \textit{nameLen}: although method name length does not directly represent computation, it may act as a weak proxy for design or implementation style.   

\begin{figure}[t]
  \centering
  \includegraphics[width=0.47\textwidth]{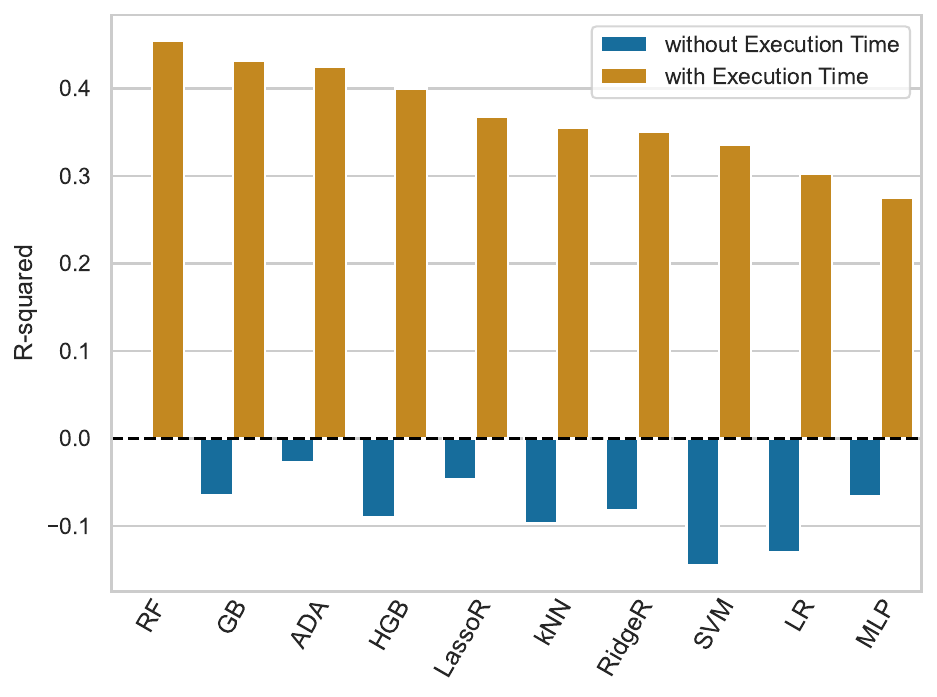}
  \caption{Effect of excluding execution time.}
  \label{fig:rq2-dropExecTime}
\end{figure}

Finally, we isolated the contribution of execution time by training the models without it. Figure~\ref{fig:rq2-dropExecTime} compares the $R^2$ values with and without execution time. Removing execution time causes a substantial performance drop across all models. For RF, $R^2$ decreases from 0.454 to 0.005 while \texttt{MAPE} increases from 1.75 to 2.27, confirming that execution time is a key lightweight dynamic proxy for predicting energy consumption at the method level.

\subsection{RQ$_3$: Hyperparameter tuning}

We tuned \texttt{RF} and \texttt{ADA} as they are the only two models featured in the top five configurations in Table~\ref{tab:top5_regression_noHyper}. Each tuned configuration retained its original feature preprocessing strategy and was evaluated using 5-fold cross-validation.

Table~\ref{tab:top5_regression_withHyper} reports the results after applying hyperparameter tuning to the five best configurations identified in $RQ_2$. For RF, three of the four configurations showed minor gains in $R^2$, with MAE and MAPE changing negligibly. \texttt{AdaBoost} slightly degraded after tuning, with $R^2$ dropping from 0.453 to 0.436 and MSE increasing from 7.218 to 7.413. MAPE improved marginally (2.152 to 2.026) but did not outweigh the overall decline. Overall, these marginal changes suggest that ensemble methods such as RF and ADA already perform reasonably well with default hyperparameters on small datasets, and that improving input representations is more impactful than model fine-tuning in this setting.
\renewcommand{\theadgape}{\Gape[0.5pt][0.5pt]}

\begin{table}[h!]
  \scriptsize
  \centering
  \caption{\textcolor{black}{Top 5 configurations based on $R^2$ score with hyperparameter tuning}}
  \setlength{\tabcolsep}{0pt}
  \begin{tabular*}{\linewidth}{@{\extracolsep{\fill}}l*{6}{r}}
    \toprule

    {\thead[l]{Model}}
    & {\thead[r]{Feature\\ Selector}}
    & {\thead[r]{Features\\ Selected}}
    & {\thead[r]{MSE}}
    & {\thead[r]{MAE}}
    & {\thead[r]{MAPE}}
    & {\thead[r]{$R^2$}}\\
    \midrule

    RF  & Variance Threshold       & 24 & 7.071 & 2.060 & 2.071 & 0.462 \\
    RF  & None                     & 35 & 7.072 & 2.060 & 2.070 & 0.462 \\
    RF  & SelectKBest              & 30 & 7.074 & 2.060 & 2.072 & 0.462 \\
    RF  & AutoSpearman             & 22 & 7.087 & 2.064 & 2.041 & 0.460 \\
    ADA & Variance Threshold       & 24 & 7.413 & 2.178 & 2.026 & 0.436 \\

    \bottomrule
  \end{tabular*}
  \label{tab:top5_regression_withHyper}
\end{table}

\begin{figure}[t]
  \centering
  \includegraphics[width=0.80\textwidth]{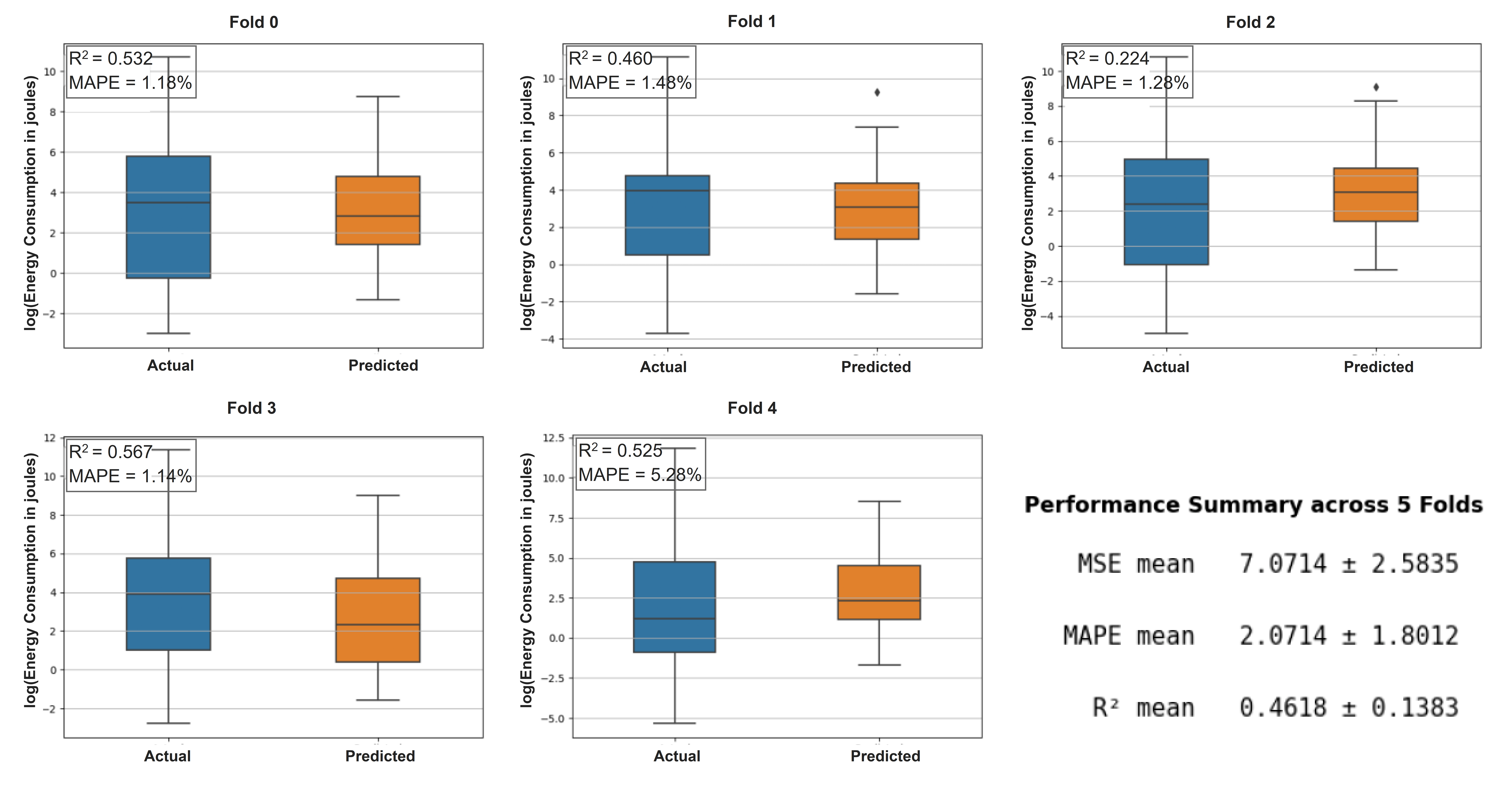}
  \caption{Box plots comparing actual and predicted log-transformed energy consumption values across all five folds for the top-performing RF configuration.}
  \label{fig:rq3-boxPlot}
\end{figure}

To further illustrate the generalization performance of the top-performing model, Figure~\ref{fig:rq3-boxPlot} presents box plots comparing actual and predicted energy values on a logarithmic scale across the five cross-validation folds. Although fold-level $R^2$ scores range from 0.224 to 0.567, the overall trend shows that predictions approximate the central tendency of the true distribution, with occasional deviation in the tails. Most folds maintain MAPE below 2\%, except for fold 4, which exhibits greater spread.

\FloatBarrier
\vspace{-0.6em}
\section{Discussion}\label{sec:discussion}
\vspace{-0.7em}

Our findings show that method-level energy consumption in Java can be partially predicted by combining static code features with execution time, though predictive performance remains modest with the best $R^2$ below 0.47. \textit{RF} emerged as the most consistent model, while simpler models such as \textit{DT} and \textit{LR} underperformed significantly, underscoring the inherent difficulty of energy estimation at fine granularity. Despite these limitations, our results help characterize how static code features relate to method-level energy consumption, providing a basis to \textit{identify energy trends rather than exact values}.

Feature selection techniques provided only slight improvements in predictive performance, but their main value lies in identifying structural code properties that correlate with energy usage, such as control flow, API usage, and method complexity. Execution time stood out as the most predictive feature across all configurations. Importantly, execution time is not included to reduce energy prediction to runtime estimation, but to quantify how much explanatory power static code features retain once runtime effects are taken into account. Prior work has shown that execution time can act as a proxy for energy consumption in mobile environments~\cite{corral2014can}. Our results empirically confirm this dominance at the method level for general-purpose Java workloads, demonstrating its significance in a domain and granularity where its impact was previously under-explored. \chg{Our results complement those of Goyal et al.~\cite{goyal2026encode}, where static features alone predict Python block-level energy with an $R^2$ of 0.75: 
there, each block is measured in isolation through amplified execution under fixed inputs to obtain a stable energy reading. The label therefore reflects the block's own computational cost rather than runtime context. Our labels instead capture method energy attributed during full-program execution, where inputs, invocation counts, and JVM runtime mechanisms dominate and are only partially captured by execution time.} Hyperparameter tuning, by contrast, delivered minimal gains, suggesting that thoughtful feature selection and preprocessing have greater impact than extensive model optimization in this setting.

Static features such as \textit{internal method calls} and \textit{cyclomatic complexity} consistently emerge as informative predictors, motivating further investigation into novel static feature based energy models and guiding developers toward refactoring efforts focused on energy optimization. Feature engineering and the incorporation of additional dynamic features such as CPU usage, which is known to significantly influence power consumption~\cite{von2016variations} represent the most promising directions for improving prediction accuracy and constitute a natural extension of this work.
\FloatBarrier
\vspace{-0.6em}
\subsection{Threats to Validity} \label{sec:energy_threats}
\vspace{-0.6em}
\noindent \textbf{External validity}.
The dataset includes \energytotalJavaMethods methods from \energytotalJavaFiles Java files, mainly drawn from Rosetta Code and CLBG benchmark tasks. These programs are benchmark-style, self-contained, and primarily algorithmic, and therefore do not fully represent large, event-driven, I/O-intensive, or framework-based software systems. The results are specific to Java programs running on our tested JVM and hardware, mainly reflecting CPU activity, and should not be directly extended to other languages, virtual machines, or architectures such as ARM servers and mobile devices. \chg{Results may further vary across JVM vendors and versions, as runtime behavior and energy attribution can differ between implementations.}

\noindent \textbf{Internal validity}.
Several factors such as just-in-time compilation, garbage collection, and operating-system scheduling can affect both time and energy measurements. To minimize their effect, all experiments were conducted on a dedicated testbed with fixed CPU frequency, disabled hyper-threading, and minimal background activity, following best practices from prior studies. Each run included warm-up and cool-down phases to reduce thermal-related measurement variation, and energy readings were averaged over repeated executions. Since multi-threaded Java implementations are not present in CLBG and Rosetta Code, all benchmarks are single-threaded, and the results therefore characterize single-threaded execution behavior only. Furthermore, filtering out incomplete and zero-energy measurements reduced our training dataset to just 265 methods, which represents a thin basis for training and comparing eleven models, limiting the reliability of the model comparison results.

\noindent \textbf{Construct validity}.
The study models method-level energy consumption using JoularJX and async-profiler. JoularJX measures power through RAPL-based domains, which cover only CPU and DRAM and omit peripheral components such as I/O, meaning methods dominated by such activity may appear as zero-energy. Its precision is further affected by hardware counter granularity, sampling interval configuration, method call frequency, and the attribution strategy used to map energy readings to individual methods. Prior work has shown that short-lived or infrequently executed methods may produce noisy or unreliable estimates~\cite{noureddine-ie-2022,burchell2023don}, and recent evaluations indicate that sampling at around 1\,ms increases profiling overhead and occasionally yields missing or zero readings, whereas longer intervals of approximately 10\,ms provide a better tradeoff between accuracy and overhead~\cite{brunnert2025evaluating}. Methods reporting zero energy were excluded from analysis. \chg{Moreover, method inputs were fixed rather than systematically varied, and code coverage was not measured during profiling, which may introduce a mismatch between static predictors and the code paths actually exercised.} These factors can introduce variance or systematic error into the measured values, potentially influencing model training and evaluation.

\noindent \textbf{Conclusion validity}.
Our use of standard metrics (\ie MAE, MSE, MAPE, and $R^2$) may not capture all aspects of predictive performance, such as sensitivity to outliers. Given the exploratory nature of this study, we do not report statistical significance testing. To mitigate these threats, we rely on multiple complementary metrics and consistent evaluation procedures across all models and configurations.

\FloatBarrier
\vspace{-0.7em}
\section{Conclusion and future work} \label{sec:energy_conclusion}
\vspace{-0.7em}
\chg{This study investigated the limits of method-level energy prediction in Java by profiling \energytotalJavaMethods Java methods to extract \energystaticFeaturesCount static features alongside execution time, and training eleven regression models to assess their predictive power.} The results show that static code metrics alone yield poor predictive performance, with $R^2$ values near zero. Adding execution time as a lightweight dynamic feature significantly improves accuracy, reaching an $R^2$ of 0.46. \textit{Execution time}, \textit{internal method calls}, and \textit{cyclomatic complexity} consistently emerge as the most influential predictors of energy consumption.

Future work could explore larger datasets, extending beyond isolated algorithmic tasks to real-world applications that utilize external frameworks. Such projects introduce new layers of abstraction and I/O operations. Evaluating these systems will require understanding how method-level energy profiles interact to impact a system's overall energy footprint. Additionally, incorporating dynamic features like CPU/GPU and memory usage could further enhance the predictive power of the models. \chg{Finally, the reusability of trained models across evolving codebases and hardware, and the energy cost of retraining them, remain open questions for future investigation.}

\vspace{-0.6em}
{\small
  \setlength{\itemsep}{0pt}
  \bibliographystyle{splncs04}
  \bibliography{references}
}

\end{document}